\documentclass[aps,prl,twocolumn,showpacs,preprintnumbers]{revtex4}
\usepackage{amsmath}
\usepackage{graphicx}
\usepackage{dcolumn}
\usepackage{bm}
\usepackage{subfigure}
\usepackage{amsfonts}
\usepackage{amssymb}

\newcommand{\qed}{\nobreak \ifvmode \relax \else
      \ifdim\lastskip<1.5em \hskip-\lastskip
      \hskip1.5em plus0em minus0.5em \fi \nobreak
      \vrule height0.75em width0.5em depth0.25em\fi}
\input{tcilatex}

\begin{document}

\title{\textbf{An exact pressure evolution equation for the incompressible
Navier-Stokes equations}}
\author{Massimo Tessarotto\thanks{%
Electronic-mail: M.Tessarotto@cmfd.univ.trieste.it}$^{1,2}$, Marco Ellero$%
^{3}$, Necdet Aslan$^{4}$, Michael Mond$^{5}$, Piero Nicolini$^{1,2,6}$ }
\affiliation{$^{1}$Department of Mathematics and Informatics, University of Trieste,
Trieste, Italy\\
$^{2}$Consortium of Magneto-fluid-dynamics, Trieste, Italy\\
$^{3}$ Institute of Aerodynamics, Technical University of Munich, Garching,
Germany\\
$^{4}$Physics Department, Yeditepe University, Kayisdagi, Istanbul, Turkey\\
$^{5}$Department of Mechanical Engineering, Ben-Gurion University of the
Negev, Beer-Sheva, Israel\\
$^{6}$INFN, Italian National Institute of Nuclear Physics, Trieste, Italy }
\date{\today }

\begin{abstract}
In this paper the issue of the determination of the fluid pressure in
incompressible fluids is addressed, with particular reference to the search
of algorithms which permit to advance in time the fluid pressure without
actually solving numerically the Poisson equation. Based on an inverse
kinetic approach recently proposed for the incompressible Navier-Stokes
equations we intend to prove that an exact evolution equation can be
obtained which advances in time self-consistently the fluid pressure. \ The
new equation is susceptible of numerical implementation in Lagrangian CFD
simulation codes.
\end{abstract}

\maketitle

An important aspect of computational fluid dynamics is related to the
determination of the fluid pressure in isothermal incompressible fluids and
particularly to the construction of an exact evolution equation for the
fluid pressure which replaces the Poisson equation. This amounts to
transform an elliptic type fluid equation into a suitable hyperbolic
equation, a result which usually is reached only by means of an asymptotic
formulation. In this paper we intend to show that an exact solution to this
problem is possible when the evolution of the fluid fields is described by
means of a suitable dynamical system, to be identified with the so-called
Navier-Stokes (N-S) dynamical system \cite{Ellero2005}. Besides being a
still unsolved mathematical problem, the issue is relevant for at least two
reasons: a) the\ proliferation of numerical algorithms in computational
fluid dynamics which reproduce the behavior of incompressible fluids only in
an asymptotic sense (see below); b) the possible verification of conjectures
involving the validity of appropriate equations of state for the fluid
pressure. Another possible motivation is, of course, the ongoing quest for
efficient numerical solution methods to be applied for the construction of \
the fluid fields\ $\left\{ \rho ,\mathbf{V,}p\right\} ,$ solutions of the
initial and boundary-value problem associated to the incompressible N-S
equations (INSE). \ For definiteness, it is convenient to recall that this
is defined by the continuity, N-S and isochoricity equations%
\begin{eqnarray}
&&\left. N\mathbf{V}=\mathbf{0},\right.   \label{INSE-2} \\
&&\left. \nabla \cdot \mathbf{V}=0,\right.   \label{INSE-4}
\end{eqnarray}%
where the mass density $\rho $ \ and the fluid pressure\textbf{\ }$p$ are
required to satisfy the inequalities 
\begin{eqnarray}
&&\left. p(\mathbf{r,}t)\geq 0,\right.   \label{INSE-5} \\
&&\left. \rho (\mathbf{r,}t\mathbb{)}=\rho _{o}>0,\right.   \label{INSE-6b}
\end{eqnarray}%
Here $N$ is the N-S operator $N\mathbf{V\equiv }\rho \frac{D}{Dt}\mathbf{V}+%
\mathbf{\nabla }p+\mathbf{f}-\mu \nabla ^{2}\mathbf{V,}$ with $\frac{D}{Dt}=%
\frac{\partial }{\partial t}+\mathbf{V\cdot \nabla }$ the convective
derivative, $\mathbf{f}$ denotes a suitably smooth volume force density
acting on the fluid element, $\rho _{o}$ is the constant mass density and $%
\mu >0$ is the constant fluid viscosity. Equations (\ref{INSE-2})-(\ref%
{INSE-6b}) are assumed to admit strong solutions in an open set $\Omega
\times I,$ with $\Omega \subseteq \mathbb{R}^{3}$ the configurations space
(defined as the subset of $\mathbb{R}^{3}$ where $\rho (\mathbf{r,}t\mathbb{)%
}>0$) and $I\subset \mathbb{R}$ a possibly bounded time interval. By
assumption $\left\{ \rho ,\mathbf{V,}p\right\} $ are continuous in the
closure $\overline{\Omega }.$ Hence if in $\Omega \times I,$ $\mathbf{f}$ is
at least $C^{(1,0)}(\Omega \times I),$ it follows necessarily that $\left\{
\rho ,\mathbf{V,}p\right\} $ must be at least $C^{(2,1)}(\Omega \times I),$
while the fluid pressure and velocity must satisfy respectively the Poisson
and energy equations%
\begin{equation}
\nabla ^{2}p=\mathbf{-\triangledown \cdot f-}\rho \nabla \mathbf{\cdot }%
\left( \mathbf{V\cdot \nabla V}\right) ,  \label{INSE-7}
\end{equation}%
\begin{equation}
\mathbf{V\cdot }N\mathbf{V}=0.  \label{INSE-8}
\end{equation}%
It is well known that the choice of the Poisson solver results important in
numerical simulations, since its efficient numerical solution depends
critically on the number of modes or mesh points used for its discretization 
\cite{Temam1983} (see also Ref.\cite{Fletcher1997} and references therein
indicated).\ In turbulent flows this number can become so large to
effectively limit the size of direct numerical simulations (DNS) \cite%
{Ellero2004}. This phenomenon may be worsened by the algorithmic complexity
of the numerical solution methods adopted for the Poisson equation. For this
reason previously several alternative approaches have been devised which
permit to advance in time the fluid pressure without actually solving
numerically the Poisson equation. Some of these methods are \emph{asymptotic,%
} i.e., to advance in time the fluid pressure they replace the exact Poisson
equation with suitable algorithms or equations which hold only in an
asymptotic sense (neglecting suitably small corrections), others are \emph{%
exact solvers,} i.e., provide in principle rigorous solutions of INSE (and
Poisson equation). The first category includes the pressure-based method
(PBM) \cite{Harlow}, the Chorin artificial compressibility method (ACM) \cite%
{Chorin1967}, the so-called preconditioning techniques \cite{Turkel99}, all
based on ACM, and kinetic approaches, of which a notable example is provided
by the so-called Lattice-Boltzmann (L-B) methods (for a review see for
example Ref.\cite{Succi} and references therein indicated). PBM is an
iterative approach and one of the most widely used for incompressible flows.
Its basic idea is to formulate a Poisson equation for pressure corrections,
and then to update the pressure and velocity fields until the isochoricity
condition (\ref{INSE-4}) is satisfied in a suitable asymptotic sense. The
ACM approach and the related preconditioning techniques, instead, are
obtained by replacing the Poisson and N-S equations with suitable
parameter-dependent evolution equations, assuming that the fluid fields
depend on a fictitious pseudo-time variable $\tau $. In dimensionless form
the evolution equation for the pressure becomes in such a case $\varepsilon
^{2}\frac{\partial }{\partial \tau }p+\nabla \cdot \mathbf{V=}0,$ where $%
\varepsilon ^{2}>0$ is an infinitesimal parameter. Manifestly this equation
recovers only asymptotically, i.e., for $\varepsilon ^{2}\rightarrow 0,$ the
exact isochoricity condition (\ref{INSE-4}). Introducing the fast variable $%
\overline{\tau }\equiv \tau /\varepsilon ^{2},$ this implies that the fluid
fields must be of the form \ $\mathbf{V}(\mathbf{r},t,\overline{\tau }),p(%
\mathbf{r},t,\overline{\tau })$ and should be assumed suitable smooth
functions of $\overline{\tau }$.$\ $Therefore, for prescribed finite values
of $\varepsilon ^{2}$ ( to be assumed suitably small), this equation permits
to obtain also an \emph{asymptotic estimate} for the fluid pressure $p(%
\mathbf{r},t)$. This is expressed by the equation%
\begin{eqnarray}
&&\left. p(\mathbf{r},t)=\lim_{\overline{\tau }\rightarrow \infty }p(\mathbf{%
r},t,\overline{\tau })\cong \right.  \\
&\cong &p(\mathbf{r},t,\overline{\tau }=0)-\int\limits_{0}^{\overline{\tau }%
^{\ast }}d\overline{\tau }^{\prime }\nabla \cdot \mathbf{V}(\mathbf{r},t,%
\overline{\tau }^{\prime }),  \notag
\end{eqnarray}%
where $\overline{\tau }^{\ast }>>1$ is suitably defined and $p(\mathbf{r},t,%
\overline{\tau }=0)$ denotes some initial estimate for the fluid pressure.
Several implementations on the Chorin algorithm are known in the literature
(see for example Refs.\cite{Housman2004,Turkel,Gaitonde,Tarnamidis}). \
Customary L-B methods are asymptotic too since they recover INSE only in an
approximate sense; moreover typically they rely on the introduction of an
equation of state for the fluid pressure, for example, the equation of state
of an ideal gas, or more refined models based on so-called non-ideal fluids 
\cite{Shi2006}. This assumption, however, generally requires that the
effective Mach-number characterizing the L-B approach, defined by the ratio $%
M^{eff}=V^{\sup }/c$ (with $c$ denoting the discretized velocity of the test
particles and $V^{\sup \text{ }}$the sup of the velocity field at time $t$%
),\ must result suitably small. As a consequence, in typical L-B approaches
the fluid pressure can only be estimated asymptotically. However, there are
other numerical approaches which in principle provide exact Poisson solvers.
These include the so-called spectral methods in which the fluid fields are
expanded in terms of suitable basis functions. Significant examples are the
pure spectral Galerkin and Fourier methods \cite{Boyd} as well as the
nonlinear Galerkin method \cite{Temam1990}, which are typically adopted for
large-scale turbulence simulations. In these methods the construction of
solution of the Poisson equation is obtained analytically. However, the
series-representation of the fluid fields makes difficult the investigation
of the qualitative properties of the solutions, such - for example - the
search of a possible equation of state or an evolution equation for the
fluid pressure.

Another approach which provides in principle an exact Poisson solver is the
one recently proposed by Ellero and Tessarotto \cite{Ellero2004,Ellero2005},
based on an \emph{inverse kinetic theory} for INSE. This approach, recently
applied also to quantum hydrodynamic equations \cite{Piero}, permits to
represent the fluid fields as moments of a suitably smooth kinetic
distribution function {$f(\mathbf{x},t)$ which obeys an appropriate inverse
Vlasov-type kinetic equation:}%
\begin{equation}
\frac{\partial }{\partial t}f+\frac{\partial }{\partial \mathbf{x}}\cdot (%
\mathbf{X}f)=0.  \label{inverse kinetic eq}
\end{equation}%
Here $\mathbf{X(x},t)\equiv \left\{ \mathbf{v,F}\right\} $ and $\mathbf{x}=(%
\mathbf{r,v)\in }\Gamma \subseteq \overline{\Omega }\times 
\mathbb{R}
^{3}$ is the state vector generated by the vector field $\mathbf{X,v}$ is
the kinetic velocity, while $\mathbf{F}(\mathbf{x,}t)$ is an appropriate
mean-field force obtained in Ref.\cite{Ellero2005}. \ In Refs. \cite%
{Tessarotto2006,Tessarotto2006b}, it has been proven that $\mathbf{F}(%
\mathbf{x,}t)$ can be uniquely prescribed, in particular, in such a way that:

\begin{itemize}
\item All the fluid equations are obtained from appropriate moments of Eq.(%
\ref{inverse kinetic eq}). As a consequence, {the fluid equations as well as
the initial and boundary conditions for the fluid fields are satisfied
identically. }

\item T{he time evolution of the kinetic distribution function, }$%
T_{t,t_{o}}f(\mathbf{x}_{o})${\ }${=f(\mathbf{x}(t),t),}$ {is determined by
the classical dynamical system associated to the vector field }$\mathbf{X,}$
i.e.,%
\begin{eqnarray}
&&\left. \frac{d}{dt}\mathbf{x}=\mathbf{X}(\mathbf{x},t)\right.
\label{N-S dynamical system} \\
&&\left. \mathbf{x(}t_{o})=\mathbf{x}_{o}\right.  \notag
\end{eqnarray}%
{\ (\emph{N-S dynamical system}) which must hold for arbitrary initial
conditions }$\mathbf{x}_{o}=(\mathbf{r}_{o}\mathbf{,v}_{o})\in \Gamma .${\ }

\item T{he solution of (\ref{N-S dynamical system}), }$\mathbf{x}%
(t)=T_{t,t_{o}}\mathbf{x}_{o},$ which defines the N-S evolution operator $%
T_{t,t_{o}},$ determines uniquely a set of curves $\left\{ \mathbf{x}%
(t)\right\} \equiv \left\{ \mathbf{x}(t),\forall t\in I\right\} _{\mathbf{x}%
_{o}}$ obtained for arbitrary $\left( \mathbf{x}_{o},t_{o}\right) \in \Gamma
\times I,$ which can be interpreted as \emph{phase-space Lagrangian
trajectories} associated to a set of fictitious \emph{"test" particles}.
Their projections onto the configuration space, denoted as \emph{%
configuration-space} \emph{Lagrangian trajectories,} are defined by the
curves $\left\{ \mathbf{r}(t)\right\} \equiv \left\{ \mathbf{r}(t)\equiv
T_{t,t_{o}}\mathbf{r}_{o},\forall t\in I\right\} _{\mathbf{x}_{o}}.$ By
varying their initial conditions, in particular $\mathbf{r}_{o}\in \Omega ,$
the curves $\left\{ \mathbf{r}(t)\right\} $ can span, by continuity, the
whole set $\overline{\Omega }.$

\item The fluid pressure $p(\mathbf{r},t)$ is defined by 
\begin{equation}
p(\mathbf{r},t)=p_{1}(\mathbf{r},t)-p_{o}(t),  \label{constitutive eq.}
\end{equation}%
(to be regarded as a{\ \emph{constitutive equation }}for{\emph{\ }} $p(%
\mathbf{r},t)$), where {$p_{1}(\mathbf{r},t)$} is the {\emph{kinetic pressure%
} $p_{1}(\mathbf{r},t)=\int dv\frac{1}{3}u^{2}f(\mathbf{x},t),$ }$p_{o}$\ is
denoted as \emph{\ reduced pressure, }while{\ }$\mathbf{u}$ is the relative
velocity $\mathbf{u}\mathbb{\equiv }\mathbf{v}-\mathbf{V}(\mathbf{r,}t).$

\item By definition, the reduced pressure $p_{o}$ is solely a function of
time, to be assumed suitably smooth and prescribed. Both $p_{o}(t)$ and {$%
p_{1}(\mathbf{r},t)$} are strictly positive, while $p_{o}(t)$ in $\overline{%
\Omega }\times I$ is subject to the constraint $p_{1}(\mathbf{r}%
,t)-p_{o}(t)\geq 0.$

\item A particular solution of the inverse kinetic equation (\ref{inverse
kinetic eq}) is provided by the local Maxwellian distribution $f_{M}(\mathbf{%
x,}t;\mathbf{V,}p_{1})=\frac{\rho _{o}}{\left( \pi \right) ^{\frac{3}{2}%
}v_{th}^{3}}\exp \left\{ -Y^{2}\right\} $ [where $Y^{2}=\frac{\mathbf{u}^{2}%
}{vth^{2}}$ and $v_{th}^{2}=2p_{1}/\rho _{o}$]. In such a case, the vector
field $\mathbf{F}$ reads:%
\begin{eqnarray}
&&\left. \mathbf{F(r,v,}t)=\mathbf{a}-\frac{1}{\rho }N_{0}\mathbf{V}+\frac{%
\mathbf{u}}{2}A_{0}p_{1}+\right.  \label{eq. for F} \\
&&\left. \frac{1}{\rho }\nabla p\left\{ \frac{\mathcal{E}}{p_{1}}-\frac{3}{2}%
\right\} \mathbf{,}\right.  \notag
\end{eqnarray}%
where $\mathbf{a}$ denotes the convective term $\mathbf{a=}\frac{1}{2}%
\mathbf{u}\cdot \nabla \mathbf{V+}\frac{1}{2}\nabla \mathbf{V\cdot u,}$ $%
\mathcal{E}$ is the relative kinetic energy density $\mathcal{E}\mathcal{=}%
\rho u^{2}/2,$ while $N_{0}$\textbf{\ }and\textbf{\ }$A_{0}$ are the
differential operators $N_{0}\mathbf{V}\equiv -\mathbf{f(r,V,}t)+\mu \nabla
^{2}\mathbf{V}$ and $A_{0}p_{1}\mathbf{(r,}t)\equiv \frac{1}{p_{1}}\left[ 
\frac{\partial }{\partial t}p_{1}+\nabla \cdot \left( \mathbf{V}p_{1}\right) %
\right] $. For an arbitrary and suitably smooth distribution function $f(%
\mathbf{x,}t),$ the form of the vector field $\mathbf{F}$ satisfying these
hypotheses has been given in Refs. \cite{Ellero2005,Tessarotto2006}.
\end{itemize}

An interesting issue is related to the consequences of the constitutive
equation (\ref{constitutive eq.}) and of the N-S dynamical system generated
by the initial value-problem (\ref{N-S dynamical system}). \ In this Letter
we intend to prove that the fluid pressure $p(\mathbf{r},t)$ obeys an exact
partial-differential equation which uniquely determines is time evolution.
This is obtained by evaluating its Lagrangian derivative along an arbitrary
configuration-space Lagrangian trajectory $\left\{ \mathbf{r}(t)\right\} $
generated by the N-S dynamical system. The result can be stated as follows.

\emph{Assuming that the initial-boundary value problem associated to INSE
admits a suitably strong solution }$\left\{ \rho ,\mathbf{V,}p\right\} $%
\emph{\ in the set }$\Omega \times I,$\emph{\ the following statements hold:}

\emph{A) If }$\mathbf{x}(t)$\emph{\ is a particular solution of} \emph{Eq.} 
\emph{\ (\ref{N-S dynamical system}) which holds for arbitrary }$\mathbf{r}%
(t)\in \Omega $ \emph{and} $t\in I,$ \emph{along each phase-space Lagrangian
trajectory }$\left\{ \mathbf{x}(t)\right\} $ \emph{defined by Eq.} \emph{\ (%
\ref{N-S dynamical system}) the} \emph{scalar field} $\xi (\mathbf{r}%
,t)\equiv $ $\mathcal{E}/p_{1}$ \emph{obeys the exact evolution equation}%
\begin{equation}
\frac{d}{dt}\xi =-\frac{1}{2}\mathbf{u\cdot }\nabla \ln p_{1}
\label{evolution equation -1}
\end{equation}%
\emph{which holds for arbitrary} \emph{\ initial conditions} $\mathbf{x}%
_{o}{}=\mathbf{(r}_{o}\mathbf{,v}_{o}\mathbf{),}$ \emph{and} $\xi _{o}=\frac{%
\rho u_{o}^{2}}{2p_{1}(\mathbf{r}_{o},t_{o})},$ \emph{with} $\mathbf{u}%
_{o}\equiv \mathbf{v}_{o}-\mathbf{V}(\mathbf{r}_{o},t_{o}).$ \emph{Here is }$%
\frac{d}{dt}$\emph{\ the Lagrangian derivative }$\frac{d}{dt}\equiv \frac{%
\partial }{\partial t}+\mathbf{v}\cdot \nabla +\mathbf{F}\cdot \frac{%
\partial }{\partial \mathbf{v}},$\emph{\ }$\xi (\mathbf{r},t),$ \emph{while
all quantities (}$\mathbf{u},E$\emph{\ and }$p_{1})$\emph{\ are evaluated
along an arbitrary phase-space trajectory \ }$\left\{ \mathbf{x}(t)\right\}
. $

\emph{B) Vice versa, if the solutions} $\mathbf{x}(t)\mathbf{=}(\mathbf{r}(t)%
\mathbf{,\mathbf{v}(}t))$ \emph{and }$\xi (t)$\emph{\ of Eqs.(\ref{N-S
dynamical system}), (\ref{evolution equation -1}) are known for arbitrary
initial conditions} $\mathbf{x}_{o}{}=\mathbf{(r}_{o}\mathbf{,v}_{o}\mathbf{%
),}$ $\mathbf{u}_{o}\equiv \mathbf{v}_{o}-\mathbf{V}(\mathbf{r}_{o},t_{o})$ 
\emph{and} $\xi _{o}=\frac{\rho u_{o}^{2}}{2p_{1}(\mathbf{r}_{o},t_{o})})$%
\emph{\ and for all }$(\mathbf{r},t)\in $\emph{\ }$\Omega \times I,$ \emph{\
it follows necessarily that in }$\Omega \times I,$ $\left\{ \rho ,\mathbf{V,}%
p\right\} $ \emph{satisfy identically INSE.}

PROOF

Let us first prove statement A), namely that INSE and the N-S dynamical
system imply necessarily the validity of \ Eq.(\ref{evolution equation -1}).
For this purpose we first notice that by construction Eq.(\ref{N-S dynamical
system}) admits a unique solution $\mathbf{x}(t)$ for arbitrary initial
conditions $\mathbf{x}_{o}=(\mathbf{r}_{o}\mathbf{,v}_{o})\in \Gamma ,$
while the same equation can also be expressed in terms of the relative
velocity $\mathbf{u\mathbf{=v-V}}(\mathbf{\mathbf{r},}t)$. This yields

\begin{equation}
\frac{d}{dt}\mathbf{u=F-}\frac{D\mathbf{V}(\mathbf{r,}t\mathbf{)}}{Dt}-%
\mathbf{u\cdot \nabla V}(\mathbf{r,}t\mathbf{)}  \label{eq-10}
\end{equation}%
Upon invoking the N-S equation (\ref{INSE-2}) and by taking the scalar
product of Eq.(\ref{eq-10}) by $\rho \mathbf{u}$, this equation implies 
\begin{equation}
\frac{d}{dt}\mathcal{E}\mathcal{=}\mathbf{u\cdot }\nabla p\left\{ \frac{%
\mathcal{E}}{p_{1}}-\frac{1}{2}\right\} +\frac{\mathcal{E}}{p_{1}}\left[ 
\frac{\partial }{\partial t}p_{1}+\nabla \cdot \left( \mathbf{V}p\right) %
\right] ,
\end{equation}%
which gives 
\begin{equation}
\frac{d}{dt}\xi \equiv \frac{\partial }{\partial t}\xi +\mathbf{v\cdot
\nabla }\xi +\mathbf{F\cdot }\frac{\partial }{\partial \mathbf{v}}\xi =-%
\frac{1}{2p_{1}}\mathbf{u\cdot }\nabla p+\mathcal{E}\nabla \cdot \mathbf{V}.
\label{evolution equation for p}
\end{equation}%
As a consequence of the isochoricity condition (\ref{INSE-4}) this equation
reduces identically (i.e., for arbitrary initial conditions for the
dynamical system) to Eq.(\ref{evolution equation -1}). \ \ B) Vice versa,
let us assume that the solutions $\mathbf{x}(t)\mathbf{=}(\mathbf{r}(t)%
\mathbf{,\mathbf{v}(}t))$ and $\xi (t)$ of Eqs.(\ref{N-S dynamical system}),
(\ref{evolution equation -1}) are known for arbitrary initial conditions $%
\mathbf{x}_{o}\in \Gamma $ and $\xi _{o}=\frac{\rho u_{o}^{2}}{2p_{1}(%
\mathbf{r}_{o},t_{o})}.$ In this case it follows the fluid fields
necessarily must satisfy INSE in the whole set $\Omega \times I.$ It
suffices, in fact, to notice that by assumption the evolution operator $%
T_{t,t_{o}}$ is known. This permits to determine uniquely the kinetic
distribution function at time $t,$ which reads \cite{Ellero2005} $f(\mathbf{x%
}(t),t)=f(\mathbf{x}_{o},t_{o})/J(\mathbf{x}(t),t),$ where $J(\mathbf{x}%
(t),t)$ is the Jacobian of the flow $\mathbf{x}_{o}\rightarrow \mathbf{x}%
(t). $ Hence, also its moments are uniquely prescribed, including both $%
\mathbf{V(r,}t)$ and \ $p(\mathbf{r,}t),$ in such a way that they result at
least $C^{(2,1)}(\Omega \times I).$ The inverse kinetic equation (\ref%
{inverse kinetic eq}), thanks to the special form of $\mathbf{F}$ as given
by Eq. (\ref{eq. for F}) ensures that the N-S equation is satisfied
identically in $\Omega \times I$ \cite{Ellero2005}. Moreover, since Eqs. (%
\ref{eq-10}) and (\ref{evolution equation for p}) are by assumption
fulfilled simultaneously, it follows that both the isochoricity condition (%
\ref{INSE-4}) and the Poisson equation [Eq.(\ref{INSE-7})] must be satisfied
too in $\Omega \times I.$ This completes the proof.

As a basic implication,\ if the fluid velocity is assumed to satisfy both
the N-S equation and isochoricity condition, the mass density satisfies the
incompressibility condition (\ref{INSE-6b}), while $\left\{ \mathbf{x}%
(t)\right\} $ is an arbitrary trajectory of the N-S dynamical system, it
follows that Eq.(\ref{evolution equation -1}) determines uniquely the
time-advancement of the fluid pressure. Hence, it provides an evolution
equation for the fluid pressure, which by definition is equivalent
simultaneously to the isochoricity condition and to the Poisson equation.
This equation can in principle be used to determine the fluid pressure at an
arbitrary position $\mathbf{r}\in \Omega .$ However, since any given
position can be reached by infinite phase-space (and also
configuration-space) Lagrangian trajectories, it is sufficient to sample the
configuration space by a suitable subset of Lagrangian trajectories (test
particles), obtained by prescribing the initial condition $\mathbf{x}_{o}$.

The physical interpretation of the pressure evolution equation is
elementary: it yields an unique prescription for the Lagrangian time
derivative of the fluid pressure, which is defined in the frame which is
locally co-moving with a test particle of state $\mathbf{x}(t)$ and velocity 
$\mathbf{\mathbf{v}(}t).$In particular, it is obvious that the specification
of the initial kinetic velocity $\mathbf{v}_{o}$ remains essentially
arbitrary, as well as the definition of the reduced pressure $p_{o}(t)$. \
This means that both the dimensionless ratios $M_{V}=V/\left\vert \mathbf{v}%
_{o}\right\vert $ and $M_{p}=p/p_{o},$ to be denoted as \emph{velocity} and 
\emph{pressure effective Mach numbers}, remain essentially arbitrary. \ As a
consequence it is possible, in principle, to construct asymptotic solutions
of Eq.(\ref{evolution equation -1}) based on low effective-Mach numbers
expansions, i.e., for which $M_{V},$ $M_{p}\ll 1$. As an illustration, let
us prove that an approximate solution of this type can be obtained for $%
p_{1} $ (and hence for $p$) in the so-called \emph{diffusive approximation,}
i.e., by considering a subset of velocity space in which by assumption at
time $t_{o}$ the initial relative velocity $\left\vert \mathbf{u}\right\vert
_{t_{o}}$ and the relative kinetic energy $\mathcal{E}$ =$\rho _{o}u^{2}/2$
are assumed to satisfy the Mach-number orderings $M_{V}\sim O(\delta )$ and $%
M_{p}\sim O(\delta ^{0}),$ being $\delta \ll 1.$ These imply $\left. \frac{%
\mathcal{E}}{p_{1}}\right\vert _{t_{o}}\sim \frac{1}{\delta },$ $\left\vert 
\mathbf{u}\right\vert _{t_{o}}\sim \frac{1}{\delta ^{1/2}}.$\ It follows
that in an infinitesimal time interval $\left[ t_{o},t_{1}=t_{o}+\Delta t%
\right] $, assuming $\Delta t\sim O(\delta ),$ there results $\frac{d}{dt}%
\mathbf{u\cong }\frac{1}{\rho }\nabla p\frac{\mathcal{E}}{p_{1}}\left[
1+O(\delta ^{1/2})\right] $ (\emph{diffusive approximation}) which yields,
by integrating it in the Euler approximation, 
\begin{equation}
\mathbf{u}(t\mathbf{)-u}(t_{o})\cong \frac{1}{\rho }\nabla p\frac{\mathcal{E}%
}{p_{1}}\Delta t\left[ 1+O(\delta ^{1/2})\right] .
\end{equation}%
In the same approximation the relative kinetic energy at time $t$ becomes 
\begin{equation}
\mathcal{E}(t)\cong \frac{\mathcal{E}(t_{o})}{1-\mathbf{u}(t_{o})\cdot
\nabla \ln p_{1}(t_{o})\Delta t}\left[ 1+O(\delta ^{1/2})\right] .
\end{equation}%
As a consequence, Eq.(\ref{evolution equation -1}) can now be used to
advance in time $p_{1}.$ In fact, integrating it and invoking again the
Euler approximation, yields%
\begin{equation}
\left. \frac{\mathcal{E}}{p_{1}}\right\vert _{t}-\left. \frac{\mathcal{E}}{%
p_{1}}\right\vert _{t_{o}}\cong -\frac{1}{2}\mathbf{u}(t_{o})\mathbf{\cdot }%
\nabla \ln p_{1}(t_{o})\Delta t,
\end{equation}%
which delivers an equation for $p_{1}(t).$ We stress that these features are
potentially important for the construction of possible numerical algorithms
based on Eq.(\ref{evolution equation -1}). Therefore, the pressure evolution
equation can in principle be adopted for the development of Lagrangian
particle simulation methods in fluid dynamics. These developments will be
the object of future investigations.

\textbf{ACKNOWLEDGEMENTS} 
Useful comments and stimulating discussions with K.R. Sreenivasan, Director,
ICTP (International Center of Theoretical Physics, Trieste, Italy) are
warmly acknowledged. Research developed in the framework of COST Action P17
"Electromagnetic Processes of Materials" [N.A., M.M. and M.T.] and PRIN
Project \textit{Fundamentals of kinetic theory and applications to fluid
dynamics, magnetofluid dynamics and quantum mechanics} (MIUR, Ministry for
University and Research, Italy), with the partial support of the ICTP
[M.E.], the Area Science Park, Trieste, Italy [P.N] and Consortium for
Magnetofluid Dynamics, Trieste, Italy. 

\end{document}